\documentclass{phyeauth}
\usepackage{graphicx}
\usepackage{amsmath}
\usepackage{amssymb}

\begin{document}

\begin{frontmatter}

\title{$1/f$ noise in a dilute GaAs two-dimensional hole system in the insulating phase}

\author[address1]{G. Deville},
\author[address2,address1]{R. Leturcq\thanksref{thank1}},
\author[address1]{D. L'H\^ote\thanksref{thank1}},
\author[address1]{R. Tourbot},
\author[address3]{C. J. Mellor}
and
\author[address3]{M. Henini}

\address[address1]{Service de Physique de l'{\'E}tat Condens{\'e}, DSM, CEA-Saclay, F-91191 Gif-sur-Yvette Cedex, France}

\address[address2]{Solid State Physics Laboratory, ETH Z{\"u}rich, CH-8093 Z{\"u}rich, Switzerland}

\address[address3]{School of Physics and Astronomy, University of Nottingham, University Park, Nottingham 
NG7 2RD, United Kingdom}

\thanks[thank1]{
Corresponding authors.\\
E-mails: lhote@drecam.saclay.cea.fr, leturcq@phys.ethz.ch}

\begin{abstract}
We have measured the resistance and the $1/f$ resistance noise of a two-dimensional low density hole system in a high mobility GaAs quantum well at low temperature. At densities lower than the metal-insulator transition one, the temperature dependence of the resistance is either power-like or simply activated. The noise decreases when the temperature or the density increase. These results contradict the standard description of independent particles in the strong localization regime. On the contrary, they agree with the percolation picture suggested by higher density results. The physical nature of the system could be a mixture of a conducting and an insulating phase. We compare our results with those of composite thin films.
\end{abstract}

\end{frontmatter}


The nature of the ground state of low density two-dimensional electron or hole systems (2DES or 2DHS) in semiconductors remains a longstanding problem. If the disorder is large, the standard description of the system is a crossover from weak localization (WL) to strong localization (SL) as its density $p_s$ is decreased \cite{Abr01}. The WL regime corresponds to Fermi liquid (FL) independent quasi-particles. Its transport properties at low temperature are well described by Drude conductivity with WL interference corrections. The transition to SL is governed by the dimensionless parameter $k_F l$ where $k_F$ is the Fermi wave vector and $l$ the mean free path. Both $k_F$ and $l$ decrease with $p_s$\cite{Keu03}, and the SL regime is reached for $k_F l < 1$. In this regime the transport is described by variable range hopping (VRH), $\rho=\rho_0 exp(T_0/T)^\gamma$ where $\rho$ is the resistivity, $T$ the temperature, $\rho_0$ a prefactor related to the hopping process, and $T_0$ a parameter depending on the localization length \cite{Shklo02}. The exponent $\gamma$ is 1/3 for a flat density of states and 1/2 for a soft Coulomb gap. 

In clean high mobility samples, the density can be decreased to such low values that the interaction effects may become large $before$ $k_F l$ reaches 1. In this case, the ratio $r_s=E_{ee}/E_c \propto m^*/p_s^{1/2}$ of the interaction to the kinetic energies reaches values much larger than 1 and the independent particle picture should break down. Correlations should appear, and for large $r_s$, the ground state for zero disorder is expected to be a Wigner crystal \cite{Tan01} or an hybrid phase \cite{Falakshahi01}. The interest for this domain had been revived by the observation of a metal-insulator transition (MIT) in two dimensions at zero magnetic field \cite{Abr03,Alt06}. The question remains on whether this new behavior is related to the transition towards a new phase where the interactions would induce strong correlations. 

Several scenarios have been proposed in what concerns the physics of low density 2DES or 2DHS in clean samples. A glassy freezing has been suggested in Si-MOSFETs \cite{Jaroszynski03,Jaroszynski04}. In GaAs 2DHS, local electrostatic studies \cite{Ilani} and transport measurements in a parallel magnetic field \cite{Gao01} suggest the coexistence of two phases. Various calculations predict a spatial separation of a low and a high density phase \cite{Spivak}. In such theories, the transport properties could be due to the percolation of the conducting phase through the insulating one \cite{Spivak}. The conducting phase could be the Fermi liquid, and the insulating one the Wigner crystal. A percolation scenario for a metal-insulator transition in two-dimensional electron or hole systems has been put forward by several authors \cite{EfrosNixon,Mei02,Meir,Shi01,DasSarma09}. Experimentally, scaling laws observed on the resistance \cite{DasSarma09,Leturcq03,Leturcq05,DasSarma08} and on the resistance fluctuations \cite{Leturcq03,Leturcq05} of 2DHS and 2DES in GaAs favor the percolation description. While the high density metallic phase has been widely studied, only few experiments have been performed in the low density insulating phase. Understanding this phase might provide useful informations to determine the microscopic nature of the transition, and the role of the correlations between electrons. In the present study, we have measured the resistance and the $1/f$ resistance noise to investigate the physical properties of the low-density phase of a 2DHS in p-GaAs.


Our 2DHS are created in Si modulation doped (311)A high mobility GaAs quantum wells. The metallic gate used to change the density is evaporated onto a 1 $\mu$m thick insulating polymide film \cite{Leturcq03,Leturcq05}. The mobility at a density $p_s = 6 \times 10^{14}$ m$^{-2}$ and a temperature $T = 100$ mK is 55 m$^2$(Vs)$^{-1}$. The experiments were carried out on Hall bars 50 $\mu$m wide, and with a distance of 300 $\mu$m between voltage probes. We performed transport and resistance noise measurements at densities ranging from $1.15\times 10^{14}$ m$^{-2}$ to $1.46\times 10^{14}$ m$^{-2}$ ($r_s > 26$) and temperatures from 50 to 800 mK. Noise measurements were made in the typical frequency interval $0.01 < f < 3$ Hz using the correlation method to reduce noise originating from the amplifiers (see Refs.~\cite{Leturcq03,Leturcq07,Deville03}). The voltage noise power spectrum $S_V$ is measured for a constant current $I$. We verified that $S_V$ is proportional to $I^2$, as expected for noise originating from resistance fluctuations. We present the normalized resistance noise power $S_R/R^2$, $R$ being the resistance of the 2DHS in the ohmic region. The noise spectra were fitted with $S_R/R^2 = A/f^{\alpha}$, $A$ being the noise magnitude at 1 Hz, and $\alpha$ an exponent which remained in the interval 0.8 to 1.5.


\begin{figure}
\begin{center}\leavevmode
\includegraphics[width=0.9\linewidth]{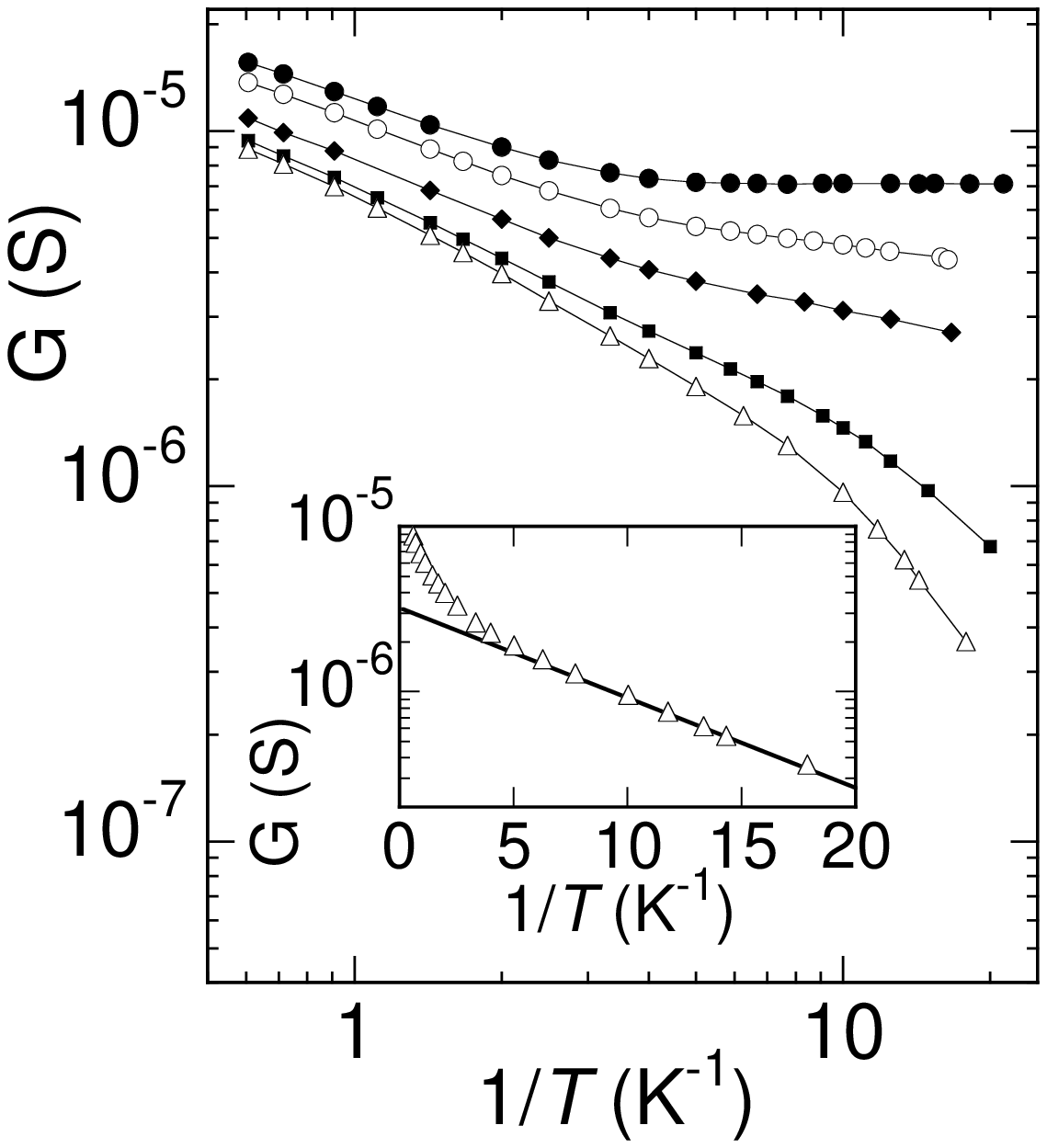}
\caption{Conductance $G$ vs. temperature $T$ for the hole densities, $p_s=1.46$, $1.42$, $1.37$, $1.33$, $1.31$ $\times 10^{14}$ m$^{-2}$ (from top to bottom). Inset: Semi-log plot of the same data for $p_s=1.31 \times 10^{14}$ m$^{-2}$. The continuous line is a fit of an activated law at low $T$.  }
\label{fig1}\end{center}\end{figure}

Figure~\ref{fig1} shows the temperature dependence of the conductance $G=1/R$ for densitites $p_s$ in the insulating phase of the MIT (which is defined by a change of the sign of $dR/dT$, and occurs at $p_s=p_c=1.46\times 10^{14}$ m$^{-2}$). A striking feature of our data is that for the highest densities, $p_l < p_s < p_c$, there is an intermediate regime where the $T$-dependence at low $T$ is {\em power}-like (open circles and closed diamonds in Fig.~\ref{fig1}). For this intermediate regime, both the low and the high $T$ parts of the curves can be fitted with power laws. On the contrary, for $p_s < p_l$ the $T$-dependence is exponential-like at low $T$.
The boundary between the two regimes corresponds to a change of the sign of the second derivative of $\log(G)$ vs. $\log(T)$ (see Fig.~\ref{fig1}). We fitted the low $p_s$ and low $T$ data with the standard variable range hopping laws \cite{Shklo02} (see above), however the best fits are obtained with a {\em simply activated} law (see inset of Fig.~\ref{fig1}). Its activation energy $T_0$ is a decreasing function of $p_s$ which goes to zero when $p_s$ goes to $p_l$. 

\begin{figure}
\begin{center}\leavevmode
\includegraphics[width=0.8\linewidth]{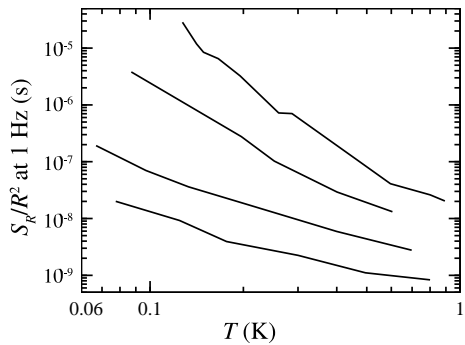}
\caption{Normalized noise power $S_R/R^2$ at 1 Hz vs. $T$ for the hole densities $p_s = 1.16, 1.23, 1.33, 1.40 \times 10^{14}$ m$^{-2}$ (from top to bottom).}
\label{fig2}\end{center}\end{figure}

\begin{figure}
\begin{center}\leavevmode
\includegraphics[width=0.9\linewidth]{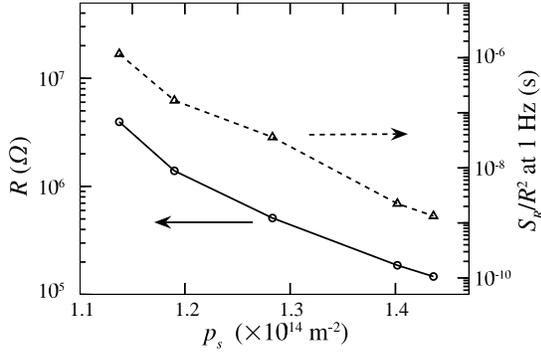}
\caption{Resistance $R$ (solid line) and normalized noise power $S_R/R^2$ at 1 Hz (dashed line) vs. the hole density $p_s$ at $T=300$ mK.}
\label{fig3}\end{center}\end{figure}

Fig.~\ref{fig2} gives the noise vs $T$ dependence. It is close to a power law, and the exponent increases in absolute value when the density decreases. Its value remains lower than $-1$. Finally we analyzed the dependence of the resistance and the noise as a function of the density. The result at a temperature $T=300$ mK is shown in Fig.~\ref{fig3}. Both $R$ and $S_R/R^2$ increase strongly when $p_s$ decreases. 

The activated $T$-dependence of the resistance we find at low temperature is not expected in the standard SL picture for which the transport is described by variable range hopping \cite{Shklo02}. This is a clear indication that a new physical understanding of our very low $p_s$ system is needed. This activated behavior can be interpreted in the percolation model  \cite{EfrosNixon,Mei02,Meir,Shi01,DasSarma08,DasSarma09,Leturcq03,Leturcq05} which was already strongly supported by our noise scaling results \cite{Leturcq03,Leturcq05}. In this model, when $p_s$ is lower than the percolation threshold $p^*$, the system is made of isolated puddles of conducting phase in the insulating one, and transport would proceed via hopping between nearest neighbor puddles. We note however that an activated behavior can also be due to a hard gap in the excitation spectrum, as expected for a Wigner crystal \cite{Pud01}.

The presence of an intermediate regime where the $R$ vs $T$ dependence is insulating ($dR/dT < 0$) but not exponential is in a qualitative agreement with the percolation model of Meir \cite{Mei02}. In this model, $R$ increases but does not diverge when $T$ goes to zero for a whole set of densities larger than the percolation threshold. Insulating linear or power laws have been observed in some systems \cite{Bog01,Jaroszynski04,MillNohLilly}. It has been argued that a $G$ vs $T$ power law with an exponent 1.5 would be due to a quantum phase transition \cite{Bog01,Jaroszynski04,Beli03}. In our case, the exponent at low temperature is lower than 1 and depends on the density. Power laws are predicted for the transport with $T$-dependent screening \cite{DasSarma} extended to low densities \cite{DasSarma05}: An energy averaging of the screening at a temperature higher than a limit $T_{\text{cross}}$ \cite{DasSarma05} leads to $\sigma \propto T$. At lower temperature, the transition to the regime where the screening effects dominate should lead to a lower exponent, as is observed in our results.
 
In what concerns the temperature dependence of the noise, we examined the possibility that it could be explained by existing models of noise for hopping transport in the SL regime. We followed the lines of Ref.~\cite{Pok01} where the noise comes from fluctuation of charges outside the Miller-Abrahams cluster. The result is a power law dependence, but the calculated exponent (-1/3 or 1/9) is larger than the values we observe (below -1) \cite{Deville03}. 

In Ref.~\cite{Leturcq03}, we stressed the fact that the scaling of the noise vs. $R$ we found suggested a percolation transition similar the one observed in metal-insulator compounds \cite{Kog04}. It is thus interesting to compare our data to those obtained in Pt/Al$_2$O$_3$ composite thin films~\cite{Mant01,Mant02}. In Ref.~\cite{Mant01} a model is presented in which metallic islands are connected to each other by one of the two components: metallic links or tunnel junctions. Decreasing the fraction $x$ of the metal leads to a percolation transition. The calculation of the global resistance shows that it is dominated by the resistance of the component with the larger fraction. This is not the case for the noise, which is always dominated by the noise of the tunnel junctions. By assuming that the noise of an individual tunnel junction does not depend on the metallic fraction $x$, the calculation shows a saturation in the noise vs. $x$ dependence below the percolation transition, in very good agreement with the data. In our case, Fig.~\ref{fig3} shows no saturation of the noise vs $p_s$ dependence at low density. We thus conclude that in the percolation description of our data, the noise of the insulating bonds (e.g. tunnel barriers between conducting puddles) should depend on the density. The changes of the puddles sizes, and thickness or height of the tunnel barriers could lead to such a density dependence.

We have measured the resistance and $1/f$ noise of two-dimensional hole systems in high mobility GaAs quantum wells. We have shown that a new physical description of the system is needed instead of the standard SL description with VRH transport. Clearly, further theoretical studies are needed to understand our measurements at low density. A percolation description which would result from intermingled conducting and insulating phases is suggested by our results.


\begin{thebibliography}{10}

\bibitem{Abr01}
E.~Abrahams, P.~W.~Anderson, D.~C.~Licciardello, and T.~V.~Ramakrishnan, Phys. Rev. Lett. 42 (1979) 673.

\bibitem{Keu03}
F.~W.~Van~Keuls, H.~Mathur, H.~W.~Jiang, and A.~J.~Dahm, Phys. Rev. B 56 (1997) 13263.

\bibitem{Shklo02}
B.~I.~Shklovskii and A.~L.~Efros, Electronic Properties of Doped Semiconductors (Springer-Verlag, Berlin, 1984).

\bibitem{Tan01}
B.~Tanatar and D.~M.~Ceperley, Phys. Rev. B 39 (1989) 5005 .

\bibitem{Falakshahi01}
H.~Falakshahi and X.~Waintal, Phys. Rev. Lett. 94 (2005) 046801.

\bibitem{Abr03}
E.~Abrahams, S.~V.~Kravchenko, and M.~P.~Sarachik, Rev. Mod. Phys. 73 (2001) 251.

\bibitem{Alt06}
B.~L.~Altshuler, D.~L.~Maslov, and V.~M.~Pudalov, Physica E 9 (2001) 209.

\bibitem{Jaroszynski03}
J.~Jaroszy\'nski, D.~Popovi\'c, and T.~M.~Klapwijk, Phys. Rev. Lett. 89 (2002) 276401.

\bibitem{Jaroszynski04}
J.~Jaroszy\'nski, D.~Popovi\'c, and T.~M.~Klapwijk, Phys. Rev. Lett. 92 (2004) 226403.

\bibitem{Ilani}
S.~Ilani, A.~Yacoby, D.~Mahalu, and H.~Shtrikman, Phys. Rev. Lett. 84 (2000) 3133; Science 292 (2001) 1354.

\bibitem{Gao01}
X.~P.~A.~Gao, A.~P.~Mills, A.~P.~Ramirez, L.~N.~Pfeiffer, and K.~W.~West, Phys. Rev. Lett. 89 (2002) 16801.

\bibitem{Spivak}
B.~Spivak, Phys. Rev. B 64 (2001) 85317; {\it ibid.} 67 (2003) 125205; B.~Spivak and S.~Kivelson, pre-print cond-mat/0310712; R.~Jamei, S.~Kivelson, and B.~Spivak, pre-print cond-mat/0408066.

\bibitem{EfrosNixon}
A.~L. Efros, Solid State Comm. 65 (1988) 1281; {\it ibid.} 70 (1989) 253; J.~A.~Nixon and J.~H.~Davies, Phys. Rev. B 41 (1990) 7929.

\bibitem{Mei02}
Y.~Meir, Phys. Rev. Lett. 83 (1999) 3506.

\bibitem{Meir}
Y.~Meir, Phys. Rev. B 61 (2000) 16470; {\it ibid.} 63 (2001) 73108.

\bibitem{Shi01}
J.~Shi and X.~C.~Xie, Phys. Rev. Lett. 88 (2002) 86401.

\bibitem{DasSarma09}
S.~Das~Sarma and E.~H.~Hwang, pre-print cond-mat/0411528.

\bibitem{Leturcq03}
R.~Leturcq, D.~L'H\^ote, R.~Tourbot, C.~J. Mellor, and M.~Henini, Phys. Rev. Lett. 90 (2003) 076402.

\bibitem{Leturcq05}
R.~Leturcq {\it et al.}, in: Proc. {SPIE}: {F}luctuations and noise in materials, volume 5469, ed. D.~Popovi\'c, M.~B.~Weissman, and Z.~A.~Racz, (Mspalomas, Spain, 2004), pp. 101--113, cond-mat/0412084.

\bibitem{DasSarma08}
S.~Das~Sarma {\it et al.}, Phys. Rev. Lett. 94 (2005) 136401.

\bibitem{Leturcq07}
R.~Leturcq, Ann. Phys. Fr. 29 (2004) 1.

\bibitem{Deville03}
G.~Deville {\it et al.}, AIP Conference Proceedings 780 (2005) 139, cond-mat/0510142.

\bibitem{Pud01}
V.~M.~Pudalov, M.~D'Iorio, S.~V.~Kravchenko, and J.~W.~Campbell, Phys. Rev. Lett. 70 (1993) 1866.

\bibitem{Bog01}
S.~Bogdanovich and D.~Popovi\'c, Phys. Rev. Lett. 88 (2002) 236401.

\bibitem{MillNohLilly}
A.~P.~Mills {\it et al.}, pre-print cond-mat/0101020; H.~Noh {\it et al.}, Phys. Rev. B 68 (2003) 165308; {\it ibid.} 68 (2003) 241308(R); M.~P.~Lilly {\it et al.}, Phys. Rev. Lett. 90 (2003) 056806.

\bibitem{Beli03}
D.~Belitz and T.~R.~Kirkpatrick, Rev. Mod. Phys. 66 (1994) 261.

\bibitem{DasSarma}
S.~Das~Sarma and E.~H.~Hwang, Phys. Rev. Lett. 83 (1999) 164; Phys. Rev. B 61 (2000) 7838(R).

\bibitem{DasSarma05}
S.~Das~Sarma and E.~H.~Hwang, Phys. Rev. B 68 (2003) 195315.

\bibitem{Pok01}
V.~Ya.~Pokrovskii, A.~K.~Savchenko, W.~R.~Tribe, and E.~H.~Linfield, Phys. Rev. B 64 (2001) 201318(R).

\bibitem{Kog04}
Sh. Kogan, {\em Electronic Noise and Fluctuations in Solids} (Cambridge University Press, 1996).

\bibitem{Mant01}
J.~V.~Mantese, W.~A.~Curtin, and W.~W.~Webb, Phys. Rev. B 33 (1986) 7897.

\bibitem{Mant02}
J.~V.~Mantese and W.~W.~Webb, Phys. Rev. Lett. 55 (1985) 2212.

\end{thebibliography}
\end{document}